# An efficient model algorithm for two-dimensional field-effect transistors

Zhao-Yi Yan,[1,2,†] Zhan Hou,[1,2,†] Fan Wu,[1,2,†] Ruiting Zhao,[1,2,†] Jianlan Yan,[1,2] Anzhi Yan,[1,2] Zhenze Wang,[1,2] Kan-Hao Xue,[3,4,*] Houfang Liu,[1,2,*] He Tian,[1,2,*] Yi Yang,[1,2,*] Tian-Ling Ren[1,2,*]

[1] *School of Integrated Circuits, Tsinghua University, Beijing 100084, China*

[2] *Beijing National Research Center for Information Science and Technology (BNRist), Tsinghua University, Beijing 100084, China*

[3] *School of Integrated Circuits, School of Optical and Electronic Information, Huazhong University of Science and Technology, Wuhan 430074, China*

[4] *Hubei Yangtze Memory Laboratories, Wuhan 430205, China*

\***Corresponding Author**, E-mail: RenTL@tsinghua.edu.cn (T.-L. Ren), yiyang@tsinghua.edu.cn (Y. Yang), tianhe88@tsinghua.edu.cn (H. Tian), houfangliu@tsinghua.edu.cn (H. Liu), xkh@hust.edu.cn (K.-H. Xue)

[†]These authors contributed equally.

## ABSTRACT

Two-dimensional materials-based field-effect transistors (2DM-FETs) exhibit both ambipolar and unipolar transport types. To physically and compactly cover both cases, we put forward a quasi-Fermi-level phase space (QFLPS) approach to model the ambipolar effect in our previous work. This work aims to further improve the QFLPS model's numerical aspect so that the model can be implanted into the standard circuit simulator. We first rigorously derive the integral-free formula for the drain-source current to achieve this goal. It is more friendly to computation than the integral form. Besides, it explicitly gives the correlation terms between the electron and hole components. Secondly, to work out the boundary values required by the new expressions, we develop a fast evaluation algorithm for the surface electrostatic potential based on the zero-temperature limit property of the 2DM-FET system. By calibrating the model with the realistic device data of black phosphorus (BP) and monolayer molybdenum disulfide (ML-$MoS_2$) FETs, the completed algorithm is tested against practical cases. The results show a typical superiority to the benchmark algorithm by two orders of magnitude in time consumption can be achieved while keeping a high accuracy with 7 to 9 significant digits.

# Introduction

Over the past 50 years, the integrated circuit industry has thrived by following the transistor's dimensional scaling strategy (also known as "Dennard scaling") [1]. However, happy scaling is nearing its end [2]. Further development in the IC industry requires co-innovation in transistor materials, processes, and functions. Beyond traditional silicon, germanium, or III-V compound semiconductors, layered two-dimensional material (2DM) semiconductors greatly enrich the research dimensions of modern transistors [3-5]. In addition to challenging the traditional silicon-based n-type and p-type field-effect transistors (FETs) with 2DMs to explore the limits of the figure of merit, 2DMs also enable ambipolar FETs to regain attention [6, 7]. During the "pre" 2DM era, such devices were mainly implemented in amorphous materials (such as amorphous silicon [8, 9]), with mobility typically around 0.1 $cm^2V^{-1}s^{-1}$, making it difficult to play a role as mainstream chips. The discovery of graphene with ultra-high ambipolar mobilities ($10^4$ $cm^2V^{-1}s^{-1}$ [10]) [11-17] opened the prelude of research on ambipolar FETs based on 2DMs. Although part of the reason is that 2DMs generally lack mature doping processes to define the polarity like silicon-based devices, the reported applications indicate that, unlike traditional chemical doping that fixes the polarity "hard" and cannot be changed, the gate-tunable conductivity mode of this kind enables devices to achieve specific functions more compactly, such as single-FET frequency doubling, nonlinear logic, etc. Therefore, introducing 2DMs is expected to inject new vitality into the development of the IC industry from the perspective of "More than Moore." Therefore, it is critical to handle both unipolar and ambipolar operating modes for the transistors to develop physical models for 2DM devices.

In principle, ambipolar transport is more challenging for modeling than the unipolar case because the latter is essentially a subset of the former. Unipolar FET's modeling has been thoroughly understood by classical works such as the Pao-Sah equation [18-21], which has been the basis for developing the BSIM models [22], now an industrial standard. However, adapting the Pao-Sah equation to the ambipolar case is not straightforward. First, the Pao-Sah equation is based on the electrostatics of MOS with a "thick" channel, consistent with its original intention to describe bulk silicon materials. The direct extension exposes the problem that the theoretical model considers an infinitely thick sample in the bulk silicon background, so the carrier areal density is essentially a divergent quantity. The well-known Fermi-potential-truncation technique was used to resolve this divergence (although it can still be easily overlooked by beginners and cause problems if not carefully studied; Sah summarized this subtle issue in his later review [23]). However, for ambipolar transport, the conduction mode switches between electron and hole types, which corresponds to the channel Fermi level scanning around the mid-band and renders the truncation technique meaningless to give a finite carrier areal density. This is the flat-band divergence difficulty inherent in the Pao-Sah model. Adopting a 2D carrier density of states can partially overcome this problem [24] because the 2D carrier density of states can naturally give the areal density without integrating over the thickness direction, thus fundamentally avoiding the divergence problem. Based on this starting point, several 2DM-FET models for unipolar cases have been reported [24-27]. The differences mainly lie in treating the carrier distribution functions and some higher-order non-ideal effects. However, the more fundamental problem involving the complete electrostatic-statistical equation for the carrier switching process in the model has yet to be seriously considered. The difficulty here is that the electrons and holes subjected to an applied external field in the semiconductor should be described by the splitted quasi-Fermi levels (QFLs), which should be explicitly shown in the equations. Solving the continuity equation is necessary to obtain the exact coupling between the QFLs for electrons and holes; thus, the drain-source current is determined. However, this is undesired by developing a compact physical model. Without detailed research on the coupling relationship, some pragmatic methods can only be adopted to overcome this problem, such as the equivalent circuit model [28], which equates ambipolar FETs to the parallel connection of n-type and p-type transistors or just ignoring the QFL splitting that should exist [29]. Although an article intentionally discussed this issue [30], the method of simply discarding the coupling term was adopted without any further argument. So, how should a complete electrostatic system with ambipolar characteristics be considered? This question has yet to be answered.

In our recent work [17], we propose the concept of QFL phase space (QFLPS) to study the coupling. The results showed that the coupling makes the ambipolar system not equivalent to the parallel connection of two independent unipolar systems. However, the good news is that this coupling can be approximated by quasi-equilibrium Fermi paths as long as the carrier recombination in the channel is strong enough. This condition should hold for typical 2DM systems of current interest. The article proves that if this condition is violated, an anomalous current hump will appear on the characteristic transfer curve, which has not yet been observed experimentally (as a note: the current humps reported in experiments [15, 31-36] occur at lower drain-source biases, which contradicts with the theoretically expected higher drain-source biases, and thus are more likely to be caused by significant gate leakage current rather than excess carriers that are not recombined). Although the final proposal seems natural and has even been assumed in previous literature [9, 29, 37], the QFLPS approach provides a systematic explanation. More importantly, it provides a phase diagram analysis method for understanding device operation modes, covering both ambipolar and unipolar devices.

Based on the progress achieved with the QFLPS approach, we further consider the challenges encountered when transferring theoretical models to actual circuit simulators, which is the problem that ultimately should be overcome for developing physical device models. Specifically, although solving differential equations can be safely omitted with the help of the QFLPS, parts of the resulting formula still need to be analytically processed. This article will focus on this aspect. More importantly, complementing pure theoretical research, we introduce experimental data to calibrate model parameters and use them to test the proposed numerical algorithm. Overall, this work aims to improve the practicality of the original QFLPS theoretical work.

## A quick review of the QFLPS model

Firstly, the main formulae of the QFLPS model are reviewed here as a necessary foundation for later discussions. The starting point of QFLPS is the drift-diffusion (DD) theory. In DD theory, the drain-source current for an ambipolar 2DM-FET can be computed by solving the equations as follows

$$j_n = -\mu_n n \nabla \varepsilon_{Fn} \tag{1}$$

$$j_p = -\mu_p p \nabla \varepsilon_{Fp} \tag{2}$$

$$\nabla j_n + R(n, p, \psi) = 0 \tag{3}$$

$$\nabla j_p - R(n, p, \psi) = 0 \tag{4}$$

$$C_{ox}(\psi - V_{GS} - V_{FB}) = -q(n - p + N_i) \tag{5}$$

$$n = \int_{E_c(\psi)}^{+\infty} D_e(\varepsilon) f(\varepsilon, \varepsilon_{Fn}) d\varepsilon \tag{6}$$

$$p = \int_{-\infty}^{E_v(\psi)} D_h(\varepsilon) f(-\varepsilon, -\varepsilon_{Fp}) d\varepsilon \tag{7}$$

where Eq. (1) and (2) represent the electron and hole current densities, Eq. (3) and (4) represent the continuity equations for electron and hole flows, and Eq. (5) is the gate-electric field equation relating the surface potential $\psi$ with electron density $n$ and hole density $p$, respectively. The flat-band voltage $V_{FB}$ and fixed-impurity charge $N_i$ are set as 0 by default to simplify the discussions. Equations (6) and (7) are the definitions for $n$ and $p$, respectively, where $D_e$ and $D_h$ are the effective electron and hole densities of states of the 2DM channel, respectively, and $f(\varepsilon, \varepsilon_F) := (1 + \exp((\varepsilon - \varepsilon_F)/kT))^{-1}$ denotes the Fermi-Dirac function with $k$ and $T$ representing the Boltzmann constant and the temperature, respectively. The two integral limits of $E_c := q\phi'_n - q\psi$ and $E_v := -q\phi'_p - q\psi$ are the functions of $\psi$, where $\phi'_n$ and $\phi'_p$ are the modified threshold voltages for electron and hole flows, respectively, distinguished from $\phi_n$ and $\phi_p$ used in the ideal model [17], which denote the Fermi potentials. The former includes the non-ideal terms that are absent in the latter. Equations (5)-(7) are called electrostatic-statistical relations (ESRs).

In essence, the QFLPS model is the integral form that is equivalent to the DD equations

$$I_{ds} = \int_{\varepsilon_{Fd}}^{\varepsilon_{Fs}} \mu_n n d\varepsilon_{Fn} + \int_{\varepsilon_{Fd}}^{\varepsilon_{Fs}} \mu_p p d\varepsilon_{Fp} \tag{8}$$

where $\varepsilon_{Fs}$ and $\varepsilon_{Fd}$ are the source and drain QFLs, respectively, and $W/L = 1$ is assumed by default. Equation (8) requires solving the integral path for $K := (q\mu_n n, q\mu_p p)^T$ on the $\varepsilon_{Fs}$-$\varepsilon_{Fd}$ plane, where $K$ is given by ESRs. In principle, this path is determined by the DD equations. Here, the QFLPS approach proves that the exact path can be safely replaced with the one located in the null-curl region of QFLPS without disturbing the final results. One typical accessible path can be

$$\varepsilon_{Fn} = \varepsilon_{Fp} \tag{9}$$

Compared with solving DD equations, the computation efficiency has been significantly improved here. Hence, ESRs Eqs (5)-(7) and Eq. (8)-(9) constitute a basic QFLPS model。

According to the review above, the remaining bottlenecks for efficiency come from two aspects: (i) integral operations required by Eq. (8) and (ii) nonlinear ESRs Eqs. (5)-(7) that have no formula solution. This article is meant to solve these two problems.

## Integral-free form for drain-source current

This section investigates how to eliminate the integral operation in Eq. (8) for $I_{DS}$. The integral-free form of $I_{DS}$ is derived term by term according to the intrinsic transport mechanism. Therefore, the ultimately derived expressions will be presented in the following discussion, and the analysis of the physical transport picture is included, which helps understand the characteristics of ambipolar systems, especially those that differ from those of unipolar systems. Although similar analytical work can be found in [27], it mainly focuses on the unipolar case, while here, we consider the more general ambipolar current.

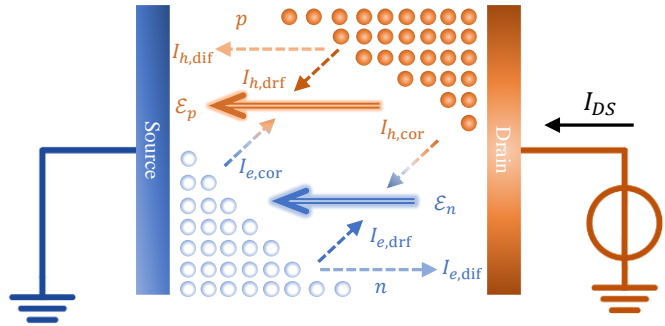

Fig. 1 Decomposition for $I_{DS}$ of ambipolar 2DM-FETs according to the QFLPS model

The total drain current can be shown (Appendix A: Current Decomposition) to be decomposed into the following six components

$$I_{DS} = I_{e,\mathrm{drf}} + I_{e,\mathrm{dif}} + I_{e,\mathrm{cor}} + I_{h,\mathrm{drf}} + I_{h,\mathrm{dif}} + I_{h,\mathrm{cor}} \tag{10}$$

where the six components on the right-hand side of the equation are introduced as follows.

(i) Electron drift current driven by the electron's built-in electric field $I_{e,\mathrm{drf}}$

$$I_{e,\mathrm{drf}} := \int_{x_s}^{x_d} q\mu_n n \mathcal{E}_n dx = \frac{1}{2} i_e \kappa_e \tilde{n}^2 |_{\tilde{n}_d}^{\tilde{n}_s} \tag{11}$$

where $\mathcal{E}_n := \frac{d}{dx}\left(-\frac{q}{C_{ox}}n\right)$ represents the electron's built-in electric field, while $x_s$ and $x_d$ denote the spatial coordinates for the source and the drain along the channel. Reduced electron density $\tilde{n} := n/kTD_e$ is defined to simplify the expression. The subscript "s" and "d" denote the source and the drain, respectively. The factor $\kappa_e := q^2 D_e/C_{ox}$ is the ratio of electron quantum capacitance to gate oxide capacitance. Electron flow coefficient $i_e := \mu_n C_{ox}(kT)^2 D_e/C_{ox}$ is introduced to collect the constant parameters arising during the derivation. This current component is shown as the deep blue dashed arrow in Fig. 1.

(ii) Electron diffusion current $I_{e,\mathrm{dif}}$

$$I_{e,\mathrm{dif}} := \int_{x_s}^{x_d} -qD_n \frac{dn}{dx} dx = i_e\{-\mathrm{Li}_2(-e^{\varphi_n})\}|_{\varphi_{n,d}}^{\varphi_{n,s}} \tag{12}$$

where $D_n = \mu_n n(dn/d(\psi - V))^{-1}$ is the electron diffusion coefficient given by the generalized Einstein relation [38, 39], $\mathrm{Li}_2(z)$ is the poly-logarithm function of order 2, and effective electron-flow driven potential $\varphi_n = (\varepsilon_{Fn} + q\psi - q\phi'_n)/kT$ is introduced, of which the values at the source and drain are denoted as $\varphi_{n,s}$ and $\varphi_{n,d}$, respectively. The electron diffusion component is expressed as the light blue dashed arrow in Fig. 1.

(iii) Electron drift current driven by hole's built-in electric field $I_{e,\text{cor}}$ (also called as the electron correlated current by us, which would be explained in following)

$$I_{e,\text{cor}} := \int_{x_s}^{x_d} q\mu_n n \mathcal{E}_p dx$$

$$= i_e \kappa_h \left\{ \text{Li}_2\left(-\frac{e^{\varphi_n}+e^{-\varphi_g}}{1-e^{-\varphi_g}}\right) + \log\left(\frac{1}{1-e^{-\varphi_g}}\right) \log\left(\frac{e^{-\varphi_g}+e^{\varphi_n}}{1-e^{-\varphi_g}}\right) - \text{Li}_2(-e^{\varphi_n}) \right\} \Bigg|_{\varphi_{n,d}}^{\varphi_{n,s}} \quad (13)$$

where the hole's built-in electric field is defined as $\mathcal{E}_p := \frac{d}{dx}\left(\frac{q}{C_{ox}}p\right)$, and factor $\kappa_h := q^2 D_h/C_{ox}$ represents the ratio of hole quantum capacitance to the gate oxide capacitance, the factor $\varphi_g := q(\phi'_n + \phi'_p)/kT$ is introduced. This component originates from the electrostatic interaction of the electrons with the hole's built-in field, as shown with the blue-orange gradient dashed arrow in Fig. 1, which manifests the correlation between the electrons and holes in the ambipolar system.

(iv) Hole drift current driven by hole's built-in electric field $I_{h,\text{drf}}$

$$I_{h,\text{drf}} := \int_{x_s}^{x_d} q\mu_p p \mathcal{E}_p dx = \frac{1}{2} i_h \kappa_h \tilde{p}^2 \Big|_{\tilde{p}_s}^{\tilde{p}_d} \quad (14)$$

where $i_h = \mu_p C_{ox}(kT)^2 D_h/C_{ox}$ is the hole flow coefficient, and $\tilde{p} := p/kTD_h$ is the reduced hole density, of which the values at the source and the drain are denoted as $p_s$ and $p_d$, respectively. $I_{h,\text{drf}}$ is depicted as the orange dashed arrow in Fig. 1.

(v) Hole diffusion current $I_{h,\text{dif}}$

$$I_{h,\text{dif}} := \int_{x_s}^{x_d} qD_p \frac{dp}{dx} dx = i_h\{-\text{Li}_2(-e^{\varphi_p})\}\Big|_{\varphi_{p,s}}^{\varphi_{p,d}} \quad (15)$$

where $\varphi_p = (-\varepsilon_{Fp} - q\psi - q\phi'_p)/kT$ is defined as the hole's effective driven potential. $I_{h,\text{dif}}$ is expressed as the light-orange dashed arrow in Fig. 1.

(vi) Hole drift current driven by electron's built-in electric field (similarly called the hole correlated current, too)

$$I_{h,\text{cor}} := \int_{x_s}^{x_d} q\mu_p p \mathcal{E}_n dx$$

$$= i_h \kappa_e \left\{ \text{Li}_2\left(-\frac{e^{\varphi_p}+e^{-\varphi_g}}{1-e^{-\varphi_g}}\right) + \log\left(\frac{1}{1-e^{-\varphi_g}}\right) \log\left(\frac{e^{-\varphi_g}+e^{\varphi_p}}{1-e^{-\varphi_g}}\right) - \text{Li}_2(-e^{\varphi_p}) \right\} \Bigg|_{\varphi_{p,s}}^{\varphi_{p,d}} \quad (16)$$

which is opposite to $I_{e,\text{cor}}$ and originates from the hole's electric field. $I_{h,\text{cor}}$ is shown as the orange-blue gradient dashed arrow in Fig. 1.

It is necessary to emphasize that the division of the "drift" or "diffusion" components mentioned here should be understood in the sense of spatial averaging, which can be seen from the integration of spatial coordinates included in the equations above.

The components derived above show the importance of considering the complete ESRs (i.e., including both electron and hole density terms) in the analysis. In ambipolar transport, the drift current of each carrier is driven by the built-in electric fields of both electrons and holes, resulting in cross terms for the drift components, i.e., $I_{e,\text{cor}}$ and $I_{h,\text{cor}}$ (Eqs. (13) and (16)). This is very different from the unipolar model reported in previous literature [25-28], where, for example, in the n-type model, the contribution of holes to the electron drift current due to the electric field interaction is usually completely ignored. This approximation is reasonable when only one type of carrier dominates. However, for cases where both electrons and holes are involved, this approximation may lead to significant differences. To give an intuitive impression of this point, we performed calculations on a practical example, and the results are shown in Fig. 2. As the gate-source voltage $V_{GS}$ gradually scans from negative to positive, the total current $I_{DS}$ firstly undergoes a decreasing and then increases, indicating a transition from holes to electrons as the dominant carrier. During this process, $I_{e,\text{drf}}$ and $I_{e,\text{dif}}$ increase with increasing $V_{GS}$, while $I_{h,\text{drf}}$ and $I_{h,\text{dif}}$ decrease with increasing $V_{GS}$. However, the trend of the correlated component of electrons and holes $(I_{e,\text{cor}} + I_{h,\text{cor}})$ is exactly opposite to this, reaching its peak in the transition interval of $V_{GS}$ because electrons and holes are most matched at this time. In other intervals where one type of carrier dominates, the coupling between electrons and holes is relatively weak due to the significant difference in their transport intensities. Therefore, for the case where ambipolar effects are important, this correlated component must be considered, requiring the consideration of the complete electrostatic equation. However, it is less emphasized in the previous related works [28-30].

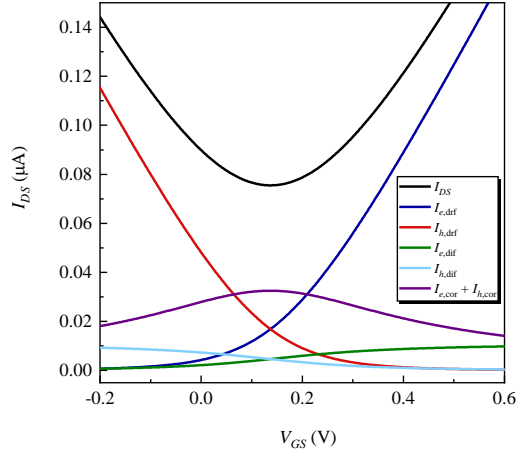

Fig. 2 Current components in the ambipolar system

Derived Eqs. (12), (13), (15), and (16) contain poly-logarithm functions, which have not been covered by Verilog A language [40]. However, there has been a standard Chebyshev rational polynomial technique to approximate it [41], which can achieve a relative error accuracy of up to $10^{-10}$ on a global scale, thus solving this problem. It is worth noting that the previous work reported poly-logarithm functions in their model, too. Nevertheless, they adopt a truncation method on its Taylor expansion [27, 30], which shows poor accuracy for the ambipolar work regime.

Given the integral-free form of $I_{DS}$ derived above, the remaining tasks are to develop a fast algorithm for those boundary values occurring in the expressions, i.e., $\tilde{n}_s$, $\tilde{n}_d$, $\tilde{p}_s$, $\tilde{p}_d$, $\varphi_{n,s}$, $\varphi_{n,d}$, $\varphi_{p,s}$, and $\varphi_{p,d}$, which, in essence, requires developing efficient solver for ESRs, i.e., Eqs. (5)-(7), which would solve the last hurdle faced in developing circuit-level models.

## Zero-Temperature Induced Solution for ESRs

An efficient algorithm for solving ESRs is going to be developed in this section. We propose to establish a functional form for the general temperature solution induced by the zero-temperature solutions of the system and then optimize the contained thermal-broadening coefficients to yield an optimal functional approximation to the exact solutions. This algorithm is inspired by the temperature scaling law satisfied by the ESRs equation. The temperature-reduced independent variables, i.e., the reduced QFLs, can be defined as

$$\tilde{\varepsilon}_{Fn} := -[-\varepsilon_{Fn} - (-q\phi'_n + qV_{GS})]/kT \tag{17}$$

$$\tilde{\varepsilon}_{Fp} := -[-\varepsilon_{Fp} - (+q\phi'_p + qV_{GS})]/kT \tag{18}$$

and the reduced dependent variable, i.e., reduced surface potential (energy), can be given as

$$\tilde{\phi} := -(q\psi - qV_{GS})/kT \tag{19}$$

With defined reduced variables, the bare ESRs can be re-written as following temperature-free form

$$F(\tilde{\phi}) := \kappa_e \ln(1 + e^{\tilde{\varepsilon}_{Fn} - \tilde{\phi}}) - \kappa_h \ln(1 + e^{\tilde{\phi} - \tilde{\varepsilon}_{Fp}}) - \tilde{\phi} = 0 \tag{20}$$

which implicitly defines the function relation between $\tilde{\phi}$, $\tilde{\varepsilon}_{Fn}$, and $\tilde{\varepsilon}_{Fp}$. The temperature $T$ becomes an implicit variable at this time. Considering the system of $T \to 0^+$ with $\tilde{\varepsilon}_{Fn}$ and $\tilde{\varepsilon}_{Fp}$ tending to $\pm\infty$, one can derive the asymptotic form for $\tilde{\phi}$ [17] as

$$\tilde{\phi}_0(\tilde{\varepsilon}_{Fn}, \tilde{\varepsilon}_{Fp}) := \frac{\kappa_e \theta_e \tilde{\varepsilon}_{Fn} + \kappa_h \theta_h \tilde{\varepsilon}_{Fp}}{\kappa_e \theta_e + \kappa_h \theta_h + 1} \tag{21}$$

which is specified as a zero-temperature limit (ZTL) formula in this work, where the two factors

$$\theta_e := \Theta\left[\tilde{\varepsilon}_{Fn} + \Lambda\left(-\frac{\kappa_h \tilde{\varepsilon}_{Fp}}{\kappa_h + 1}\right)\right] \tag{22}$$

$$\theta_h := \Theta\left[-\tilde{\varepsilon}_{Fp} + \Lambda\left(\frac{\kappa_e \tilde{\varepsilon}_{Fn}}{\kappa_e + 1}\right)\right] \tag{23}$$

are the electron and hole phase indicators, respectively. According to the phase of QFLs, they take the value of 0 or 1. Specifically, $\theta_e = 1$ if and only if the wealthy-electron phase is assumed, whereas $\theta_h = 1$ if and only if the wealthy-hole phase is in place [17].

$\forall \varepsilon_{Fp}$ and $\varepsilon_{Fp}$, constructed ZTL formula of Eq. (21) fulfills the limit of

$$\lim_{T \to 0^+} (\tilde{\phi} - \tilde{\phi}_0) = 0 \tag{24}$$

However, a general temperature solution for $\tilde{\phi}$ is desired. Inspired by the fact that the Fermi-Dirac distribution function degenerates into a step function at zero temperature, we reverse this process. Then we can obtain a functional form for the finite-temperature solution. Performing the following formal substitutions

$$\Theta(x) \to \frac{1}{1 + \exp(t^{-1}x)} \tag{25}$$

$$\Lambda(x) \to s \ln(1 + e^{s^{-1}x}) \tag{26}$$

where $t$ and $s$ are the introduced scaling parameters, one arrives at a phase indicator function that can be adapted to finite-temperature cases

$$\Theta_{s,t,\kappa}(x, y) = \frac{1}{1 + \exp\left\{-t^{-1}\left[x + s \ln\left(1 + \exp\left[-\frac{\kappa y}{(\kappa+1)s}\right]\right)\right]\right\}} \tag{27}$$

Defining

$$\Theta_e = \Theta_{s_e, t_e, \kappa_h}(\tilde{\varepsilon}_{Fn}, \tilde{\varepsilon}_{Fp}) \tag{28}$$

$$\Theta_h = \Theta_{s_h, t_h, \kappa_e}(-\tilde{\varepsilon}_{Fp}, -\tilde{\varepsilon}_{Fn}) \tag{29}$$

leads to the tentative solution function as

$$\tilde{\phi}_1(s)(\tilde{\varepsilon}_{Fn}, \tilde{\varepsilon}_{Fp}) = \frac{\kappa_e \Theta_e \tilde{\varepsilon}_{Fn} + \kappa_h \Theta_h \tilde{\varepsilon}_{Fp}}{\kappa_e \Theta_e + \kappa_h \Theta_h + 1} \tag{30}$$

where the coefficients contain the following four parameters

$$s = [s_e, t_e, s_h, t_h] \tag{31}$$

remaining to be optimized. In principle, it is required to solve a 2D plane optimizing problem as

$$s_{2D} = \underset{s'}{\mathrm{argmin}} \sum_{\tilde{\varepsilon}_{Fn}, \tilde{\varepsilon}_{Fp}} \|\tilde{\phi}_1(s')(\tilde{\varepsilon}_{Fn}, \tilde{\varepsilon}_{Fp}) - \tilde{\phi}(\tilde{\varepsilon}_{Fn}, \tilde{\varepsilon}_{Fp})\| \tag{32}$$

However, it can be worked around. Since the QFLPS model only uses the solution of $\tilde{\phi}$ along the path defined by Eq. (9), i.e., $\{(\varepsilon_{Fn}, \varepsilon_{Fp}) \mid \varepsilon_{Fn} = \varepsilon_{Fp}\}$, or written in the reduced-symbol as

$$\tilde{\varepsilon}_{Fp} - \tilde{\varepsilon}_{Fn} = \varphi_g \tag{33}$$

a better option is to solve this path-wise problem

$$\boldsymbol{s}(\varphi_g) = \underset{\boldsymbol{s}'}{\text{argmin}} \sum_{\tilde{\varepsilon}_{Fn}} \left\| \tilde{\phi}_1(\boldsymbol{s}')(\tilde{\varepsilon}_{Fn}, \tilde{\varepsilon}_{Fp}) - \tilde{\phi}(\tilde{\varepsilon}_{Fn}, \tilde{\varepsilon}_{Fp}) \right\| \tag{34}$$

where $\varphi_g$ is allowed to vary in a proper range, corresponding to a set of possible paths on QFLPS. Direct calculation of Eq. (34) is inconvenient due to the exact $\tilde{\phi}$ requiring numerical computing. A reasonable scheme is to convert the original problem defined by Eq. (34) into the form of the residue of $F(\tilde{\phi}_1)$. The extracted function relation $\boldsymbol{s}(\varphi_g)$ is not available in a compact form/ Hence, a linear regression step is needed

$$\mathcal{L} = \underset{\mathcal{L}'}{\text{argmin}} \left\| \mathcal{L}'(\varphi_g) - \boldsymbol{s}(\varphi_g) \right\| \tag{35}$$

where $\mathcal{L}$ denotes the approximated linear mapping operator and can be represented by a real-number matrix. Albeit some accuracy is missing after linear regression, this can be patched by performing several steps of newton iterations as

$$\tilde{\phi}^{(i+1)} = \tilde{\phi}^{(i)} - F'(\tilde{\phi}^{(i)})^{-1} F(\tilde{\phi}^{(i)}), \text{for } 0 \leq i \leq X - 1 \tag{36}$$

where $\tilde{\phi}^{(0)} = \tilde{\phi}_1[\mathcal{L}(\varphi_g)](\tilde{\varepsilon}_{Fn}, \tilde{\varepsilon}_{Fp})$.

## Summary of Algorithm

The derived formulae of Eqs. (10)-(36) lead to a complete QFLPS model algorithm, which comprises three modules (Fig. 3): **modular i** initial seed generator (defined by Eqs. (34) and (35)), **modular ii** iterative potential solver (Eq. (36)), and **modular iii** main current model (Eqs. (10)-(16)). Among them, **modular i** requires solving once to match the currently adopted device process, which yields a linear operator $\mathcal{L}$. And then, **modular ii** can online yield iterative initial based on $\mathcal{L}$ to compute the solution of $\tilde{\phi}$ in response to the $\tilde{\varepsilon}_{Fn}$ and $\tilde{\varepsilon}_{Fp}$ input by **modular iii**. Fed back with $\tilde{\phi}$, the **modular iii** calculates the boundary values $\tilde{n}_s, \tilde{n}_d, \tilde{p}_s, \tilde{p}_d, \varphi_{n,s}, \varphi_{n,d}, \varphi_{p,s}$, and $\varphi_{p,d}$, which are input into Eqs. (10)-(16) to finally obtain $I_{DS}$.

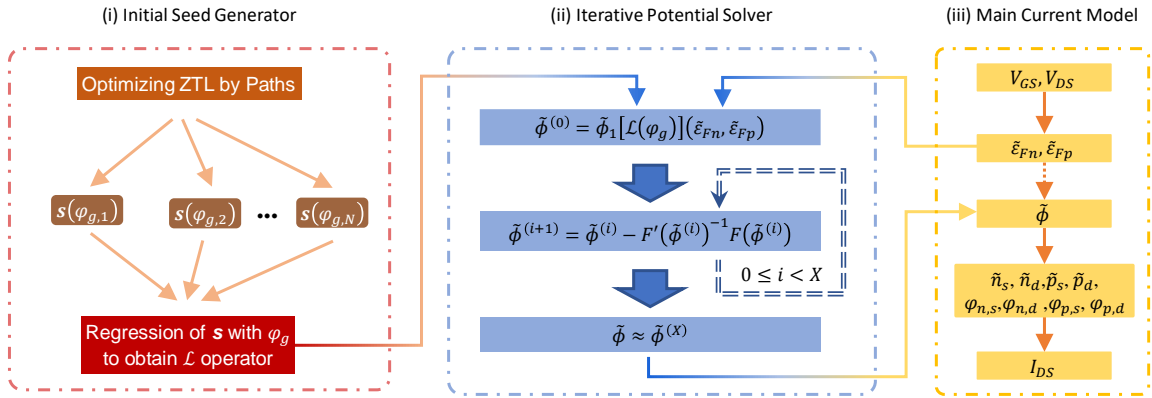

Fig. 3 Algorithm flow chart

To demonstrate the above algorithm, we exemplify it with experimental cases as follows.

## Experiment demonstrations

We select few-layer black phosphorus (FL-BP) and monolayer molybdenum disulfide (ML-MoS$_2$) as researching samples since the former represents a typical ambipolar and the latter gives a unipolar (n-type) transport properties when prepared as a FET device. The unipolar version of the QFLPS model can be readily obtained by discarding the hole-related terms. First, the QFLPS model is benchmarked with the experimental data from the fabricated FET devices to extract the model parameters. Then the developed algorithm can be applied with the corrected model to test its accuracy and speed. The benchmark algorithm uses MATLAB's built-in functions to calculate the numerical integrals and solve nonlinear ESR equations. The extraction flow and the extracted parameters are presented in Appendix B: Parameter Extraction and Appendix C: Extracted Model Parameters, respectively. The respective cases of BP and MoS$_2$ are introduced below.

## BP-FET

BP is the first verified 2D elementary semiconductor that has a finite bandgap. It has a tunable direct bandgap ranging from 0.3 eV in bulk to 2.0 eV in monolayer, which gives it a natural advantage in optoelectronic applications.

Here, we focus on its ambipolar characteristics. Due to its vulnerability to moist air, the freshness of the BP layer affects the transport properties when used in the FET device. Given this fact, we adopt a buried gate structure to reduce the number of process steps, and the overall process flow is summarized in Fig. 4a. Detailed steps can be found in the "Device Fabrication Processes." The optical microscope image of the fabricated device is shown in Fig. 4b. Raman spectrum of the channel region exhibits clear and sharp characteristic peaks of BP (Fig. 4c), indicating the high quality of the prepared film. Based on the BP-FETs prepared, ambipolar transport characteristics can be measured.

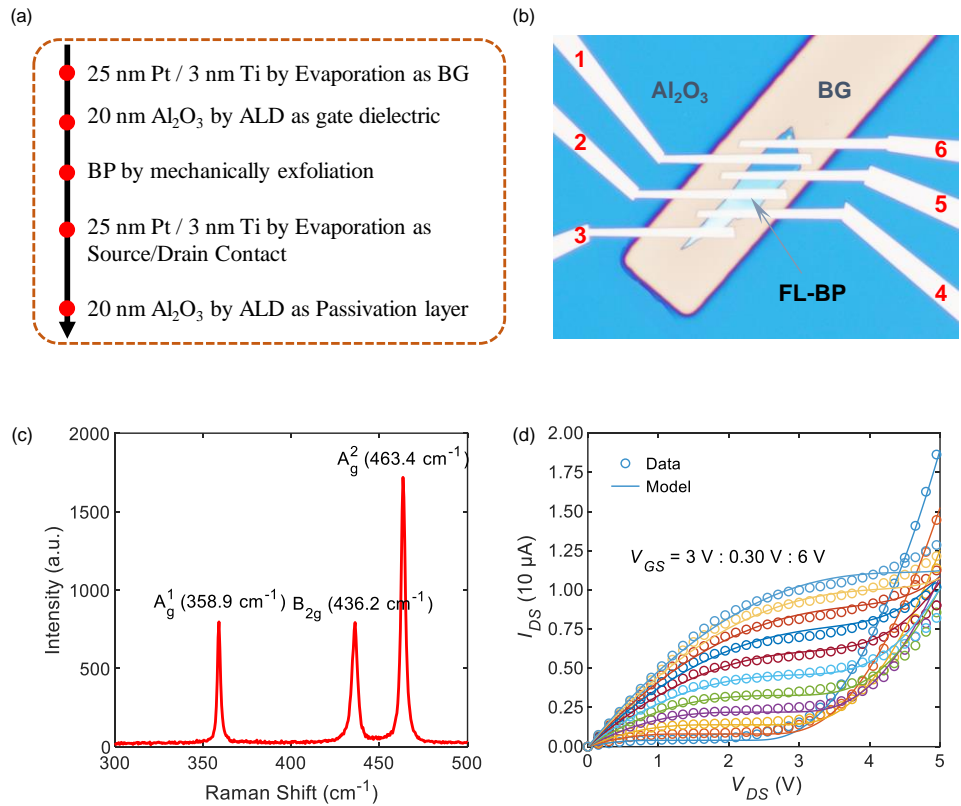

Fig. 4 Fabricated BP-FET devices. (a) process flow, (b) optical microscopic graph, (c) Raman spectrum, (d) output characteristic data and the model simulations.

The BP-FET defined by probes 1 and 3 is studied to demonstrate the model algorithm. Its output characteristic curve is shown in Fig. 4d with circle data, which gradually transitions from onset hole current to saturated electron current as the gate voltage $V_{GS}$ increases, consistent with theoretical predictions [17]. After the model parameters were obtained via the calibration procedure, the model simulation results matched well with the experimental data (solid line in Fig. 4d). Based on the parameters obtained from the experiment, the algorithm flow is demonstrated below.

First, following the **modular i** in Fig. 3, one generates the initial seed. According to the extracted parameters, sweeping $\varphi_g \in [0,30]$ is sufficient for the optimizing problem defined by Eq. (34). The optimized $s$ vector is shown as the asterisk data points in Fig. 5; Its linear regression approximation is plotted as the dashed line in Fig. 5, of which the equation coefficients comprise the matrix representation for the linear operator $\mathcal{L}$ shown as the inserted table in Fig. 5.

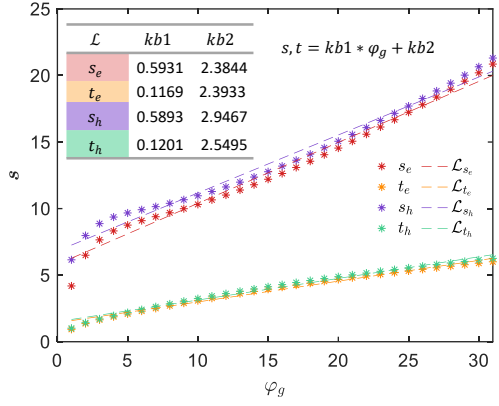

Fig. 5 Tentative solution's optimized *s* parameters and its linear regression approximation with $\varphi_g$

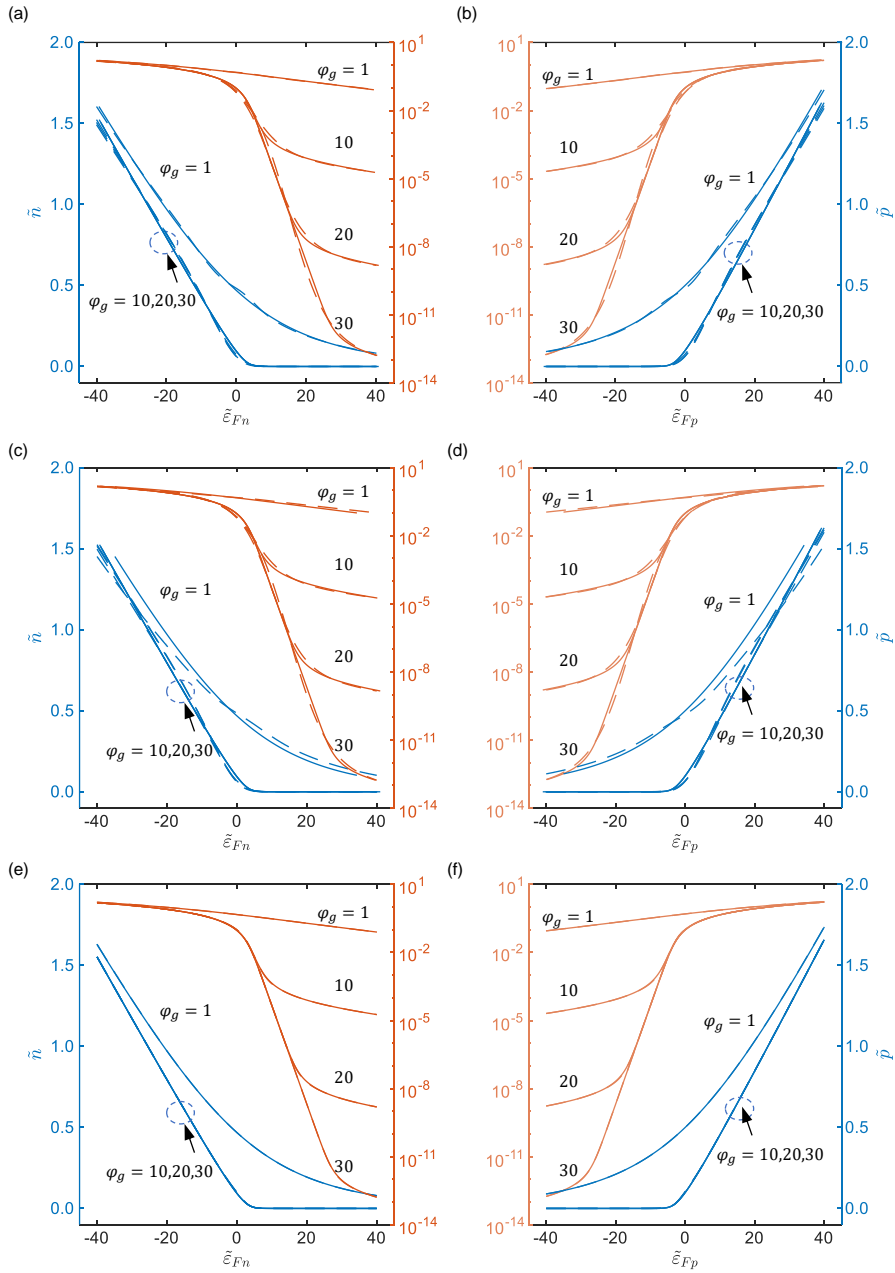

Fig. 6 Step analysis for ESRs iterative algorithm, where the algorithm predicted carrier's densities (dashed) and its exact solution (solid) are compared step by step. It uses carrier density as the comparison object than the potential itself since the former is positive-definite and thus can be presented appropriately with the logarithm scale. The dashed lines in (a-b), (c-d), and (e-f) are generated by the functions $\tilde{\phi}_1[s(\varphi_g)]$, $\tilde{\phi}_1[\mathcal{L}(\varphi_g)]$, and $\tilde{\phi}_1[\mathcal{L}(\varphi_g)]$'s one-shot Newton iteration, respectively.

To show the necessities of workflow, we check the solution accuracy achieved by the above steps in Fig. 6. The optimized $s$ gives a fair closed solution $\tilde{\phi}_1[s(\varphi_g)](\tilde{\varepsilon}_{Fn}, \tilde{\varepsilon}_{Fp})$ to the answer that the reduced electron and hole densities are sufficiently closed to the exact solutions (Fig. 6a and b). After making linear substitutions with $\mathcal{L}$, the precision drops slightly (Fig. 6c and d) to obtain a trackable form. A one-shot Newton iteration is performed so that the algorithm and the exact solution agree again (Fig. 6e and f).

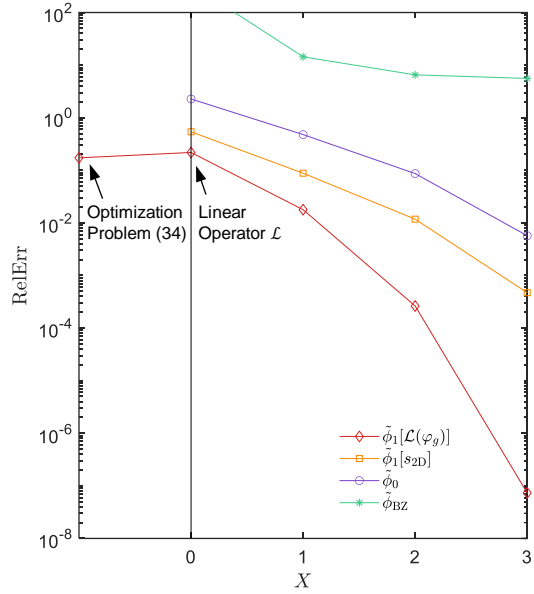

Fig. 7 Relative error (RelErr) iterative curves for $\tilde{\phi}$ from all kinds of algorithms

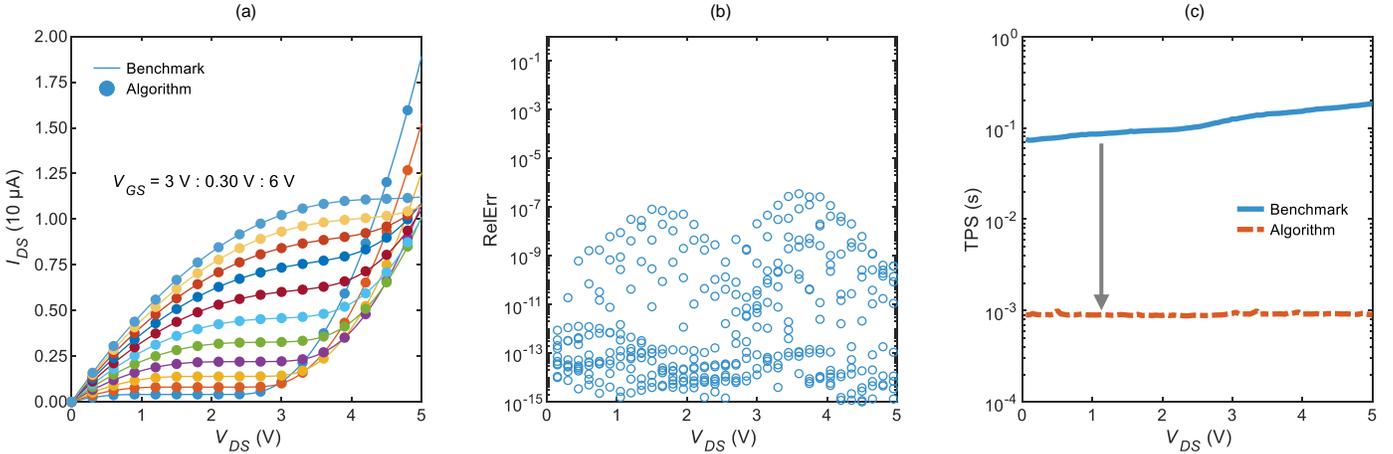

Fig. 8 Completed model test. (a) comparison of simulated output characteristics, where dots are for the proposed algorithm, while the lines are for the benchmark results; (b) the relative error calculated from the results in (a); (c) comparison of time-per-shot (TPS); $X = 3$ is adopted in the Newton iterations.

As a comparison, other initial generation schemes are studied. Besides the main algorithm defined by Eqs. (34) and (35), other methods include: (a) $\tilde{\phi}_1[s_{2D}]$, where $s_{2D}$ is given by solving Eq. (32); (b) ZTL $\tilde{\phi}_0$, given by Eq. (21); (c) the Boltzmann approximation-based initial reported by [29], which can be expressed with the symbols of our work as

$$\tilde{\phi}_{BZ}(\tilde{\varepsilon}_{Fn}, \tilde{\varepsilon}_{Fp}) = \kappa_e e^{\tilde{\varepsilon}_{Fn}} - \kappa_h e^{-\tilde{\varepsilon}_{Fp}} \tag{37}$$

The resulting convergence processes are presented in Fig. 7. It is shown that $\tilde{\phi}_1[\mathcal{L}(\varphi_g)]$ leads to as low as $10^{-8}$ relative error after a three-step Newton iteration, showing the second-order convergence speed and superior to others. On the contrary, the Boltzmann approximation scheme possesses the lowest convergence rate due to the limited applicable regime determined by its underlying subthreshold assumption. If adopting $\tilde{\phi}_1[s_{2D}]$ as the initial, which makes it free from the second regression operation that is required by the $\tilde{\phi}_1[\mathcal{L}(\varphi_g)]$, the generated iterative curve slows down accordingly since the quality of the initial is impaired by the over-simplified parameters; Hence, it might not be strange that the case becomes worse if ZTL formula is directly used, where no optimization is employed.

Given the comparison above, it is safe to conclude that the conclusion that proposed $\tilde{\phi}_1[\mathcal{L}(\varphi_g)]$ offers a compact and practical form to approximate the exact solution.

Finally, the algorithm simulations are compared with the benchmark results, as shown in Fig. 8a, which exhibit excellent agreement with each other; The relative error is as low as $10^{-7}$ (Fig. 8b); What is more important, our algorithm outperforms the benchmark in the time efficiency by two orders of magnitude(Fig. 8c).

## MoS$_2$-FET

MoS$_2$ has a stable unipolar transport characteristic and can be synthesized by the chemical vapor deposition (CVD) method on a large scale. The superior electrostatic tunability of monolayer (ML) MoS$_2$ enables the ultra-short gate length FET device [42] down to sub-1nm. Here, we prepare the ML-MoS$_2$ transistor (Fig. 9a) with CVD and the buried-gate process. The ML thickness can be confirmed by its feature peak of $E_{2g}^1$ at 386.4 cm$^{-1}$ in its Raman spectrum [43], as shown in Fig. 9b. The measured output curves show an n-type characteristic (Fig. 9c), which gets a more substantial saturation current level as $V_{GS}$ grows.

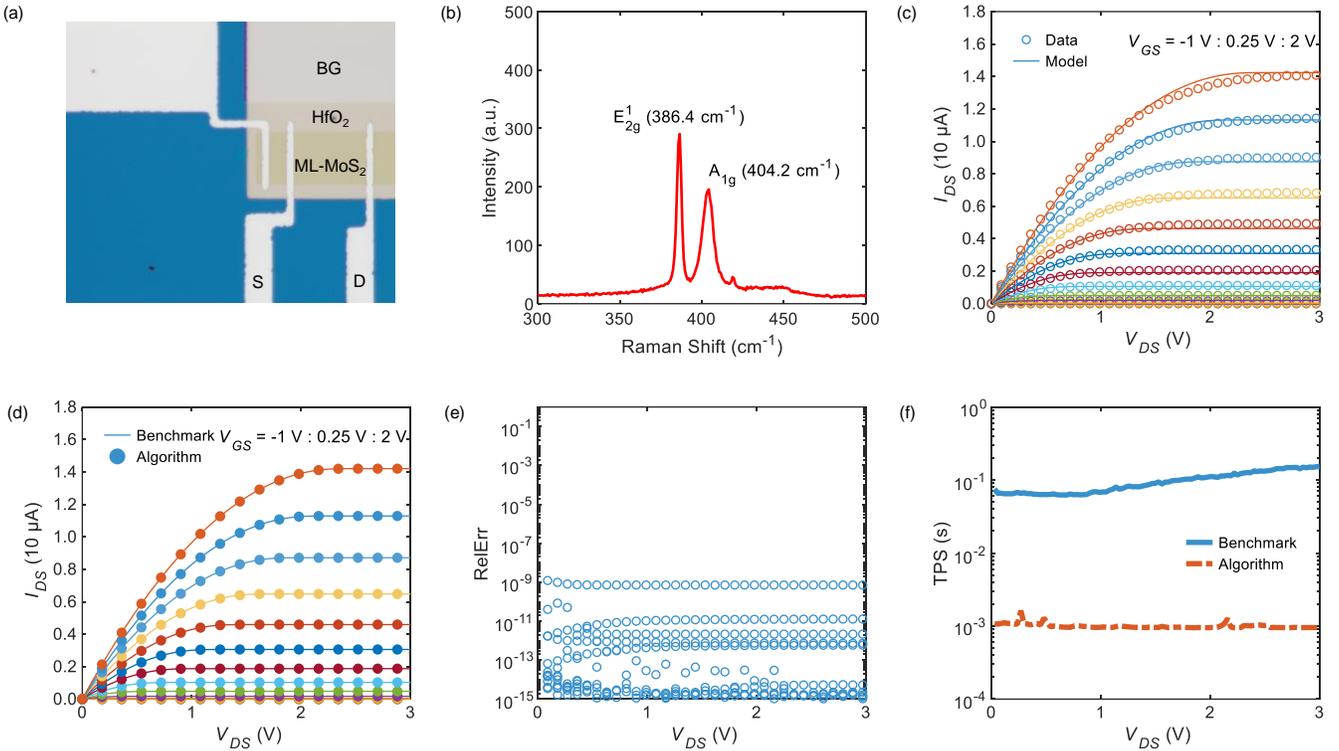

Fig. 9 Algorithm benchmark with ML-MoS$_2$-FET. (a) Device optical microscopic graph; (b) Raman spectrum; (c) Output curves for measured data (circles) and model simulations (lines); (d) Simulation results by the proposed algorithm (dots) and the benchmark results (lines); (e) Relative error calculated for (d); (f) Time-per-shot comparison.

As with BP-FET, we extract the model parameters for the fabricated ML-MoS$_2$ FET device. The simulation results given by the calibrated model are shown as the lines in Fig. 9c, which yields fairly matched results with the experimental data. For the MoS$_2$-FET case, the algorithm test for the accuracy (Fig. 9d and e) and computational efficiency (Fig. 9f) shows consistently high quality with those achieved for BP-FET.

## Conclusion

In summary, this work is the numerical algorithm extension of our previous work on the QFLPS approach. QFLPS approach provides a compact theoretical description of the 2DM-FET operation, and an integral formula for $I_{DS}$ is prescribed there. To convert it into a circuit-simulator deployable form, we develop an efficient algorithm for it and verify the algorithm with experimental instances, including BP and MoS₂. The results demonstrate the acceleration ability of the algorithm with typical two orders of magnitude while keeping a high precision (relative error $10^{-7} \sim 10^{-9}$ as for the tested cases here). The reported results present a handy tool mapping the proper device characteristic into the 2DM-based circuit design platform to make the system-level design possible.

## Device Fabrication Processes

### BP-FET

First, electron beam evaporation (EBV) was used to deposit 25 nm Pt / 3 nm Ti onto the 300 nm SiO₂ substrate as the buried gate metal stack. Then, 20 nm Al₂O₃ is deposited as the gate dielectric layer by atomic layer deposition (ALD); Later, FL-BP flakes are mechanically exfoliated and transferred onto the buried gated region of the Al₂O₃ layer. Next, electron-beam lithography (EBL) is performed to pattern the drain/source electrodes, followed by 25 nm Pt / 3 nm Ti deposition. At last, a 20 nm Al₂O₃ passivation layer is grown by ALD to protect the BP channel.

### MoS₂-FET

CVD monolayer MoS₂ is provided by Shenzhen Sixcarbon technology Co. LTD. 50 nm Ti/Pd was first deposited by EBV as local gate metal. 9 nm HfO₂ was then deposited by the Finland Picosun ALD system. For the MoS₂ wet transfer, the PMMA layer with 400K molecular weight was first spin-coated at 3,000 rpm for 60 s on Si/SiO₂/MoS₂. After coating the PMMA layer, 3 wt% potassium hydroxide solution at 110 °C was used to lift off the MoS₂/PMMA from the Si/SiO₂ substrate. Then, the MoS₂/PMMA was transferred to fresh deionized water six times to remove the potassium hydroxide residue. The cleaned MoS₂/PMMA stack was transferred to the target sample, and the excess deionized water was air-dried naturally for more than 12 h. Further annealing at 85 °C for 30 min enhanced the adhesion between the MoS₂ and the substrate. The PMMA was removed by soaking for 30 min with fresh acetone, which was done at least twice. After patterned by direct laser scribed, O₂ plasma is used to etch extra MoS₂. 50 nm Al is deposited as contact metal.

## Appendix A: Current Decomposition

This appendix derives the drain-source current's components. From Eq. (8), the electron and hole parts can be derived, respectively. For instance, the electron part can be sorted as

$$I_e = \int_{\varepsilon_{Fd}}^{\varepsilon_{Fs}} \mu_n n d\varepsilon_{Fn} = q\mu_n \left[ \int_{V_s}^{V_d} n d\psi + \int_{V_d}^{V_s} n d(\psi - V) \right] \tag{A1}$$

where the QFL voltage equivalent $V := -\varepsilon_{Fn}/q$ is introduced here as the dummy variable for integration, while $V_d$ and $V_s$ denote the QFLs at the drain and source, respectively. The subscripts "d" and "s" represent the "drain" and "source," respectively. The first integral in the bracket of Eq. (A1), which reads

$$I'_{e,\text{drf}} = \mu_n \int_{x_s}^{x_d} n \frac{d\psi}{dx} dx = \mu_n \int_{V_s}^{V_d} n d\psi \tag{A2}$$

can be recognized as the (spatially-averaged) drift component for electrons, and the second integral

$$I_{e,\text{dif}} = \int_{x_d}^{x_s} D_n \frac{dn}{dx} dx = \int_{V_d}^{V_s} D_n dn = \int_{V_d}^{V_s} \mu_n n \left( \frac{dn}{d(\psi-V)} \right)^{-1} dn = \mu_n \int_{V_d}^{V_s} n d(\psi - V) \tag{A3}$$

can be identified as the (spatially-averaged) diffusion component for electrons. It is worth noting that the generalized Einstein relations $D_n = \mu_n n (dn/d(\psi - V))^{-1}$ [38, 39] have been used. Eq. (A3) requires no further simplification since $n$ has already been the explicit function of $\psi - V$ according to Eq. (6). On the contrary, Eq. (A2) for $I'_{e,\text{drf}}$ should be transformed further. Hence, by using the differential identity

$$d\psi = \frac{\partial \psi}{\partial n} dn + \frac{\partial \psi}{\partial p} dp \tag{A4}$$

and noting that the partial derivatives in Eq. (A4) can be given by Eq. (5), namely $\partial \psi / \partial n = q/C_{ox}$ and $\partial \psi / \partial p = -q/C_{ox}$, one can sort Eq. (A2) for $I'_{e,\text{drf}}$ as

$$I'_{e,\mathrm{drf}} = \mu_n \int_{\psi_s}^{\psi_d} n d\psi = \mu_n \frac{q}{C_{ox}} \int_{n_d}^{n_s} n dn + \mu_n \frac{q}{C_{ox}} \int_{p_s}^{p_d} n dp \tag{A5}$$

The derivation shows that the first integral of Eq. (A5) represents the electron drift current driven by the electron's built-in electric field, i.e.,

$$I_{e,\mathrm{drf}} \coloneqq \mu_n \frac{q}{C_{ox}} \int_{n_d}^{n_s} n dn \tag{A6}$$

while the second one represents the electron drift current driven by the hole's built-in electric field, which requires a small amount of further treatment, i.e.,

$$I_{e,\mathrm{cor}} \equiv \mu_n \frac{q}{C_{ox}} \int_{p_s}^{p_d} n dp = \mu_n \frac{q}{C_{ox}} \int_{V_s-\psi_s}^{V_d-\psi_d} n \frac{dp}{d(V-\psi)} d(V-\psi) \tag{A7}$$

where the integrand $ndp/d(V-\psi)$ has already been the explicit function of $(V-\psi)$, thus Eq. (13) being obtained.

The derivation for the hole component is similar to that of the electron case and will not be elaborated on here.

## Appendix B: Parameter Extraction Flow

The linear approximation of the QFLPS approach [17] shows that the model recovers the drain-source current's well-known linear and quadratic dependences with the bias. In the on-state region for electrons, the electron density $n$ can be approximated by

$$n \approx (V_{GS} - \phi'_n - V)C_{ox} \tag{B1}$$

If biased to be unsaturated state ($V_{DS} < V_{GS} - \phi'_n$), the current can be approximated as

$$I_{DS} = \frac{W}{L} \int_0^{V_{DS}} q\mu_n n dV \approx \mu_n C_{ox} \frac{W}{L} \left[ (V_{GS} - \phi'_n)V_{DS} - \frac{1}{2}V_{DS}^2 \right] \tag{B2}$$

For a small $V_{DS}$, it can be further evaluated as

$$I_{DS} \approx \mu_n C_{ox} \frac{W}{L} (V_{GS} - \phi'_n) V_{DS} \tag{B3}$$

according to which one can extract the mobilities $\mu_n$ and the threshold voltage $\phi'_n$ for electron.

For the hole's parameter, the onset of hole current under high $V_{DS}$ ($V_{DS} > V_{GS} + \phi'_p$, where $V_{GS} < \phi'_n$) can be used to extract the parameters. By linear approximation, one again has

$$I_{DS}(V_{GS}, V_{DS}) \approx \mu_p C_{ox} \frac{W}{L} \left[ \frac{1}{2} V_{DS}^2 - (V_{GS} + \phi'_p) V_{DS} \right] \tag{B4}$$

Hence, the hole's mobility $\mu_p$ can be determined by the linear transconductance $G_{tp}$ with $V_{DS}$

$$G_{tp} \equiv \frac{I_{DS}(V_{GS1}, V_{DS}) - I_{DS}(V_{GS2}, V_{DS})}{V_{GS2} - V_{GS1}} \approx \mu_p C_{ox} \frac{W}{L} V_{DS} \tag{B5}$$

And the hole's threshold voltage $\phi'_p$ can be determined from the relation below

$$S_{tp} \equiv \frac{1}{2} V_{DS}^2 - I_{DS} / \left( \mu_p C_{ox} \frac{W}{L} \right) = (V_{GS} + \phi'_p) V_{DS} \tag{B6}$$

The electron's and hole's relative effective masses for the effective density of states ($D_{e(h)} = g_s g_v m^*_{e(h)} m_0 / \pi \hbar^2$) can be obtained in the literature[44, 45], where $m^*_e = 0.15$ and $m^*_h = 0.14$ are set for BP, while $m^*_e = 0.55$ and $m^*_h = 0.56$ are set for MoS$_2$.

The intrinsic QFLPS model uses an ideal 2D electrostatics, where the gate voltage changes the channel potential near a ratio of 1:1, making the model predicted subthreshold swing (*SS*) close to the Boltzmann limit. However, the realistic system possesses non-ideal factors such as interface state, which renders the *SS* of devices staying away from the limit. Hence, it is necessary to introduce an *SS*-correction factor $\eta$ [21, 46] to the bare temperature $T$ (so it becomes $\eta T$), which is allowed to be dependent on $V_{GS}$ here and can be modeled by a Gaussian-type function as

$$\eta = \eta_0 + H \exp\left[-\beta^{-2}(V_{GS} - V_{GS,0})^2\right] \tag{B7}$$

where $\eta_0$ denotes the baseline for the correction factor, H represents the varying amplitude, $\beta$ denotes the feature voltage, and $V_{GS,0}$ is the reference gate voltage.

## Appendix C: Extracted Model Parameters

Table CI Experimental parameters and the model parameters

| | Quantity | Case I | Case II |
|---|---|---|---|
| IV-data | Polarity | Ambipolar | n-type |
| Experiment parameters | Channel material | BP | MoS$_2$ |
| | Preparation process | ME | CVD |
| | Gate oxide material | Al$_2$O$_3$ | HfO$_2$ |
| | Gate oxide thickness (nm) | 20 | 9 |
| | Channel width (μm) | 8.4 | 26 |
| | Channel length (μm) | 8.5 | 20 |
| | Channel thickness (nm) | 20 | 0.72 |
| QFLPS intrinsic parameter | $m_e^*$ | 0.15 | 0.55 |
| | $m_h^*$ | 0.14 | 0.56 |
| | $\mu_e$ (cm$^2$/Vs) | 3.64 | 3.05 |
| | $\mu_h$ (cm$^2$/Vs) | 16.4 | \ |
| | $\phi_n'$ (V) | 2.42 | −0.28 |
| | $\phi_p'$ (V) | −0.27 | 2.08 |
| Subthreshold correction | H | 7.35 | 2.40 |
| | $V_{GS,0}$ | 5.20 | −0.77 |
| | $\beta$ | 1.24 | 1.09 |
| | $\eta_0$ | 2.54 | 1.31 |